\newif\ifanonymized
\anonymizedfalse    

\ifanonymized
    
\else
    
\fi

\newif\ifsubmit
 \submittrue	

\documentclass[sigconf]{acmart}

\usepackage{fancyhdr}

\usepackage{url}

\usepackage{color}

\usepackage[mmddyyyy,hhmmss]{datetime}

\usepackage{xspace}

\usepackage{graphicx}
\DeclareGraphicsExtensions{.pdf,.png,.eps,.ps,.jpg}

\usepackage{makecell}

\usepackage{multirow}
\usepackage{multicol}

\ifsubmit
  \newcommand{\hey}[1]{\relax}
  \newcommand{\heyvarun}[1]{\relax}
  \newcommand{\bibtex}[1]{\relax}
  \newcommand{\note}[1]{\relax}
\else
  \newcommand{\hey}[1]{\textcolor{magenta}{[{#1}]}}
  \newcommand{\heyvarun}[1]{\textcolor{blue}{[Varun: {#1}]}}
  \newcommand{\note}[1]{\par\textcolor{magenta}{Note: {#1}}\par}
  \newcommand{\bibtex}[1]{\textcolor{red}{@bibtex}\{#1\}} 
\fi

\newcommand{\hide}[1]{\relax}

\newcommand{\figlabel}[1]{\label{fig:#1}}
\newcommand{\figref}[1]{Figure~\ref{fig:#1}}

\newcommand{\addfigure}[4]{
\begin{figure}[tbp]
\centerline{\resizebox{#2\linewidth}{!}{\includegraphics{figs/#1}}}
\caption{\figlabel{#1}#3}
\Description{#4}
\end{figure}
}

\newcommand{\tablabel}[1]{\label{tab:#1}}
\newcommand{\tabref}[1]{Table~\ref{tab:#1}}

\newcolumntype{L}[1]{>{\raggedright\arraybackslash\hyphenpenalty=10000\exhyphenpenalty=10000}p{#1}}

\newcommand{\seclabel}[1]{\label{sec:#1}}
\newcommand{\secref}[1]{Section~\ref{sec:#1}}

\begin{document}

\title[Supporting Couples' Social Well-being in Daily Life]{Supporting Couples' Social Well-being in Daily Life: A Needs Assessment and Co-design with Co-located Couples}
\date{}

\ifanonymized
    \author{Anonymized for blind submission}
\else
    \author{Yuna Naito}
    \email{watanabe.y@northeastern.edu}
    \orcid{0000-0001-7530-8788}
    \affiliation{
     \institution{Northeastern University}
     \city{Boston}
     \state{MA}
     \country{USA}}
    \author{Timothy Bickmore}
    \email{t.bickmore@northeastern.edu}
    \orcid{0000-0002-2676-9788}
    \affiliation{
     \institution{Northeastern University}
     \city{Boston}
     \state{MA}
     \country{USA}}
    \author{Varun Mishra}
    \email{v.mishra@northeastern.edu}
    \orcid{0000-0003-3891-5460}
    \affiliation{
     \institution{Northeastern University}
     \city{Boston}
     \state{MA}
     \country{USA}}
    \author{Matthew S. Goodwin}
    \email{m.goodwin@northeastern.edu}
    \orcid{0000-0002-4237-601X}
    \affiliation{
     \institution{Northeastern University}
     \city{Boston}
     \state{MA}
     \country{USA}}
\fi




\begin{abstract}
Romantic relationships shape physical, mental, and social well-being, and relationship quality is built through everyday interactions. Addressing minor issues early and reinforcing positive interactions help maintain relationship quality, and ubiquitous computing creates new opportunities to sense and support couples' relationships in daily life. However, existing systems remain largely developed with limited involvement from couples, raising questions about which types of sensing and support couples may accept in their private lives. To address this gap, we conducted exploratory needs-assessment interviews and co-design sessions with five co-located couples (10 participants) to identify their needs and preferences for relationship-attunement-enhancing technologies, including the use of LLMs. Couples' own designs converged on systems that detect meaningful moments---such as conflict onset---and offer subtle mediation rather than AI-generated advice, highlighting dyadic concerns around partners' agency, authenticity, and control over shared interaction data. We offer preliminary design implications for ubiquitous sensing and intervention systems that center the couple as a unit rather than the individual.
\end{abstract}



\copyrightyear{2026}
\acmYear{2026}
\setcopyright{cc}
\setcctype{by}
\acmConference[UbiComp Companion '26]{Companion of the 2026 ACM International Joint Conference on Pervasive and Ubiquitous Computing}{October 11--15, 2026}{Shanghai, China}
\acmBooktitle{Companion of the 2026 ACM International Joint Conference on Pervasive and Ubiquitous Computing (UbiComp Companion '26), October 11--15, 2026, Shanghai, China}
\acmDOI{10.1145/3798063.3841773}
\acmISBN{979-8-4007-2533-3/2026/10}

  
 \maketitle  



\ifsubmit
    \relax
\else
    \par\noindent \textcolor{red}{\textbf{DRAFT}: \today\ -- \currenttime}
    \pagestyle{fancy}
    \lhead{DRAFT in preparation}
    \rhead{version: \today\ -- \currenttime}
    \chead{}
    \lfoot{}
    \rfoot{}
\fi


\section{Introduction} 
\seclabel{introduction}
Well-being encompasses not only physical and mental health but also social wellness, and the quality of intimate relationships strongly shapes all three: higher relationship satisfaction is linked to better physical health outcomes~\cite{kiecolt-glaser_lovesick_2017}, while relationship distress lowers both partners' daily well-being. Importantly, relationship well-being is built in daily life. Marital disaffection develops gradually~\cite{kayser_when_1993} through the accumulation of everyday interactions~\cite{kirchler_love_2001}. Addressing minor issues early and reinforcing positive aspects to offset negative ones~\cite{van_steenbergen_how_2021} is essential to maintaining a healthy relationship. Yet many couples seek support only after problems escalate, often in clinical settings and after years of delay~\cite{doherty_how_2021}. Ubiquitous computing has the potential to close this gap by enabling relationship support in daily life.

While conceptually promising, ubiquitous computing research on well-being has largely focused on the individual, such as sensing personal stress or mood and delivering just-in-time support. Focusing on the couple as a unit for sensing and intervention remains a nascent area of research. Prior work with couples has focused on enhancing a sense of relatedness for long-distance couples, conveying a partner's presence, and sharing personal cues~\cite{branham_expanding_2012, branham_designing_2013}, with less attention to co-located couples' everyday interactions. Advances in ubiquitous and mobile sensing make it possible to measure relationship dynamics in daily life, e.g., from conversational tone or text exchanges, and to intervene at the right moment. The emergence of Large Language Models (LLMs) further expands this design space toward systems that can analyze couples' everyday communication and support reflection and repair. In fact, some couples have already begun using LLMs to identify unhealthy patterns and enhance perspective-taking~\cite{alqurashi_impact_2025}, and emerging work uses LLMs to support communication in close relationships~\cite{wang_futuring_2026, jiang_scaffolded_2026, zhang_two_2026, fan_is_2026}. However, these systems are largely developed with limited involvement from couples in the design process. Bringing sensing and LLM-based support into couples' private lives raises important questions: what do couples want, what sensing will they accept, and where are their boundaries of privacy and autonomy? These questions should be addressed before such systems are built and deployed.

To investigate these questions, we use \textit{attunement}, defined as ``the ability to be aware of one's own state of mind and body while also tuning in and connecting to another person''~\cite{brodkin_missing_2021}, as a lens for examining how ubiquitous technology might help \textit{co-located} couples (i.e., couples living together or in close proximity) maintain and restore relationship well-being in daily life. Attunement is a useful lens because it comprises learnable, interactive skills that manifest in observable moment-to-moment behavior, the same everyday moments in which partners notice, interpret, and respond to each other. Although attunement overlaps with related constructs, including communication skills, empathy, and perceived partner responsiveness, it is not reducible to any one of them. Instead, attunement integrates these elements into a moment-to-moment dyadic process that can reveal concrete opportunities for technological support.

In this work, we conduct exploratory needs-assessment interviews and co-design sessions with \textit{co-located} couples to investigate the following research questions: (\textbf{RQ1}) What support do couples want in their daily lives to enhance attunement? (\textbf{RQ2}) What are couples' preferences for attunement-enhancing technology in daily life, including in-situ sensing and the use of LLMs? Our findings offer preliminary guidance for developing ubiquitous computing systems that sense and support couples' social well-being in daily life, identifying which forms of everyday sensing, data sharing, and technological support couples will accept.



\section{Related work} 
\seclabel{related}
Over the past two decades, HCI researchers have explored digital interventions to enhance relatedness (e.g., connectedness, intimacy, and love), with many focusing on long-distance couples. These interventions often provide abstracted presence or deep interpersonal sharing to foster relatedness across distance~\cite{branham_expanding_2012, branham_designing_2013}. Examples include a bed that simulates a partner’s presence through heat and movement~\cite{dodge_bed_1997}, smartwatches that enable partners to share and respond to each other’s biosignals~\cite{liu_significant_2021}, and ambient displays that convey emotions and presence~\cite{li_emotional_2020}. Hassenzahl et al.~\cite{hassenzahl_all_2012} analyzed this body of work and identified six common strategies for mediating relatedness: awareness, expressivity, physicality, gift-giving, joint action, and memory. Although these systems primarily target long-distance couples, co-located couples may also benefit from mediated interaction~\cite{branham_designing_2013}. 

Recent work exploring technologies to support co-located couples includes that by Griggio et al.~\cite{griggio_augmenting_2019}, who developed a mobile app that shares location data to help co-located couples better understand each other, and Wagener et al.~\cite{wagener_togetherreflect_2025}, who introduced a Virtual Reality (VR) application for collaborative drawing to reduce negative affect and support perspective-taking. Other systems explicitly target communication: Jiang et al.~\cite{jiang_remini_2025, jiang_scaffolded_2026} designed an LLM-powered chatbot that scaffolds reciprocal self-disclosure to improve positive affect, bonding, and perceived responsiveness. Chun et al.~\cite{chun_conflictlens_2025} built an LLM-based system that analyzes conflict interactions and offers feedback, which participants found helpful for managing conflict. Emerging work also uses LLMs to support individual partners, for example, by assisting with drafting text messages~\cite{fan_is_2026} or by deploying an agent that represents each partner's point of view and negotiates with the other partner's agent~\cite{zhang_two_2026}.

Although these emerging technologies have revealed some user preferences, much of this work remains technology-driven rather than grounded in broad assessments of couples' needs. Silina et al.~\cite{silina_love_2018} directly examined users' preferences among Hassenzahl et al.'s six relatedness strategies and found that high-concept designs do not always align with end users' desires, underscoring the need for user-centered rather than top-down technology design. More recently, Wang et al.~\cite{wang_futuring_2026} conducted a user study to uncover opportunities, challenges, and preferences around AI-mediated social life. While some of their insights extend to couples, their study addressed social relationships broadly; romantic couples, with their unique intimacy, interdependence, and privacy concerns, may present distinct needs and boundaries that such general studies cannot fully capture. We sought to address this gap by conducting exploratory needs-assessment interviews and co-design activities with co-located couples, with a particular focus on enhancing attunement, given its critical role in relationship maintenance.

\section{Methods} 
\seclabel{methods}

To uncover technological needs and preferences for enhancing relationship attunement, we conducted sessions with co-located couples. Each session consists of (1) needs-assessment interviews, in which we ask about potential areas for improvement in their attunement, and (2) co-design activities, in which we ask couples to ideate on a technology they would like to use in their daily lives. Needs-assessment interviews empirically surface couples’ values, and co-design sessions begin translating these values into technological concepts. The study was approved by the institutional ethics committee at Northeastern University.

\subsection{Participants}
\seclabel{participants}
We recruited five couples, with at least one partner enrolled at Northeastern University. Recruitment was conducted through online posts, flyers, and word of mouth. Inclusion criteria were as follows: age 18 or older, English fluency, a romantic relationship of at least three months, the ability to participate in the study with their partner, and residence in the greater Boston area. Exclusion criteria were self-reported current or past clinical mental health diagnoses and any history of domestic violence by past or current partners. Each participant received a \$30 gift card upon completion.

\subsection{Questionnaires}
\seclabel{questionnaire}
Before each session, we asked participants to complete questionnaires individually. To quantify relationship satisfaction, we used the Couple's Satisfaction Index (CSI)~\cite{funk_testing_2007}, a validated 32-item questionnaire with a 6-point Likert scale. We also used a modified questionnaire~\cite{brodkin_missing_2021} to measure couples' attunement. Although not yet validated, we used this questionnaire to inform our interview questions (\secref{interview}) because of its face validity and conceptual consistency with the literature on couples' satisfaction and therapy. 
The attunement questionnaire is organized around four components---relaxed awareness (i.e., self and partner awareness), listening (i.e., attentive listening), understanding (i.e., emotional understanding), and mutual responsiveness (i.e., responsive turn-taking). We adapted this framing to create parallel forms in which each participant rates both themselves and their partner on the same items. 


\subsection{Needs assessment interviews}
\seclabel{interview}
 We conducted 30-minute semi-structured needs-assessment interviews using an exploratory elicitation technique we call the ``dyadic discrepancy probe,'' which leverages discrepancies between partners’ self and each other’s perceptions to surface latent needs and opportunities for support. Couples may not always find it easy to articulate their needs at the level of small, everyday attunement behaviors. Prior work shows that congruence between self- and partner-rated empathy is associated with relationship satisfaction~\cite{kimmes_self_2014}. We therefore hypothesized that gaps between self- and partner-rated attunement indicate unmet needs that open-ended questions alone may miss, and that probing these gaps helps identify areas for improvement. More specifically, using couples' responses to the attunement questionnaire, we identified gaps between how they see themselves and how their partners see them. We computed score differences for each item, summed the scores for each of the four components, and identified the component with the largest deviation. We did not disclose their answers but provided probing questions based on the identified gap, such as what understanding looks like to them, when it is more challenging to demonstrate, and how technology could help them improve those skills. 

\subsection{Co-design activities}
\seclabel{co-design}
At the start of the co-design session, we presented three example technology ideas in text and solicited immediate reactions. This step helps participants envision the design space and articulate preferences: (1) An LLM-based tool that analyzes text conversations and suggests ways to strengthen communication; (2) An earbud-based tool that monitors in-person conversational tone and offers gentle prompts when the tone becomes less constructive; (3) A gamified VR environment that supports playful attunement exercises. We selected these examples to span a range of technologies, settings, target interactions, and intervention timing, as well as varying degrees of ubiquity, from an on-demand tool to always-on ambient sensing. For each concept, we held a semi-structured discussion to capture first impressions and the reasons behind them. Participants then designed their ideal technology to further enhance their relationships, based on the areas for improvement identified during the interview. To structure ideation, we provided a card deck introducing seven ``axes'' for thinking about technology design: time, target interaction, setting, interface, technology, input data, and content---drawing on prior card-based design toolkits~\cite{mora_tiles_2017}. The seven axes were researcher-generated sensitizing dimensions, informed by HCI design-space methods and context-aware computing’s emphasis on who, where, when, what activity/context is sensed, and how systems respond. They were intended to scaffold ideation rather than an exhaustive taxonomy. Participants were advised to select one or more cards for each axis that aligned with their preferences and to sketch or write down their ideal technology to enhance their relationship. We then conducted a brief semi-structured interview about their thought processes and preferences. The co-design session lasted approximately 60 minutes.

\subsection{Analysis}
\seclabel{analysis}
One researcher transcribed the audio and conducted a thematic analysis in NVivo~\cite{lumivero_nvivo_2025}. Coding focused on perceived challenges to couples' attunement, contexts and constraints for technology use, and preferred properties of attunement-enhancing technology. Co-design artifacts (i.e., card selections and sketches) were coded to identify patterns and derive design guidelines. We consolidated the codes into broader themes, which all authors reviewed and refined.

\section{Results} 
\seclabel{results}

\subsection{Participants}
\seclabel{results-participants}
We present participants' demographics in \tabref{participant-demographics}. We conducted two in-person sessions and three online sessions.

\begin{table*}[tbp]
\caption{\tablabel{participant-demographics}Participants' demographics. Avg. CSI = average Couples Satisfaction Index score (couple-average); Rel. length = relationship length; Married = marital status; Cohab. = cohabiting (living together).}
\centerline{\resizebox{1.0\linewidth}{!}{\begin{tabular}{lccccll}
\toprule
\textbf{Couple} &
\textbf{Avg. CSI} &
\textbf{Rel. length} &
\textbf{Married} &
\textbf{Cohab.} &
\textbf{Partner 1} &
\textbf{Partner 2} \\
\midrule

c01 & 99.5 & > 3 yrs & Y & Y &
\makecell[l]{p01: 25-34, Male, Asian/Pacific Islander} &
\makecell[l]{p02: 25-34, Female, Asian/Pacific Islander} \\

c02 & 153.5 & > 3 yrs & N & Y &
\makecell[l]{p03: 18-25, Female, White/Caucasian} &
\makecell[l]{p04: 18-25, Male, White/Caucasian} \\

c03 & 145.0 & 2-3 yrs & N & N &
\makecell[l]{p05: 18-25, Male, Black and White} &
\makecell[l]{p06: 18-25, Female, White/Caucasian} \\

c04 & 145.0 & 6-12 mos & N & N &
\makecell[l]{p07: 18-25, Female, Asian/Pacific Islander} &
\makecell[l]{p08: 18-25, Male, Asian/Pacific Islander} \\

c05 & 135.0 & 2-3 yrs & N & N &
\makecell[l]{p09: 18-25, Female, Asian/Pacific Islander} &
\makecell[l]{p10: 18-25, Male, Asian/Pacific Islander} \\

\bottomrule
\end{tabular}
}}
\end{table*}


\subsection{What support do couples want in their daily lives to enhance attunement? (\textbf{RQ1})}
\seclabel{results-rq1}

\addfigure{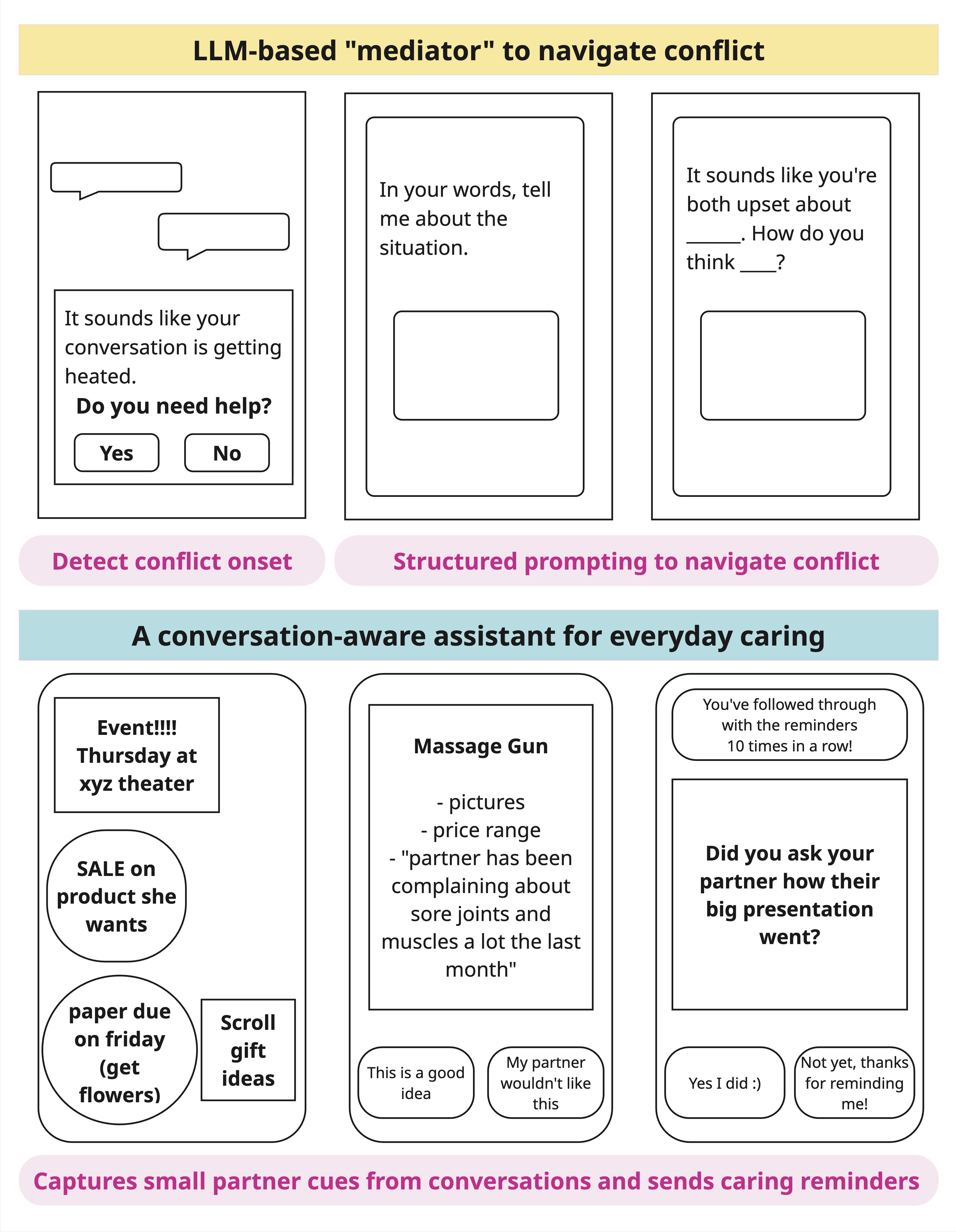}{0.85}{Example ideas generated by participants.\protect\footnotemark}{Two participant-generated interface concepts. The top concept is an LLM-based mediator for conflict: the system detects when a conversation becomes heated, asks whether the couple wants help, and then guides them through structured prompts to describe the situation and reflect on each other’s perspectives. The bottom concept is a conversation-aware assistant for everyday caring: it captures small cues about a partner’s needs or upcoming events from conversations, surfaces relevant reminders or gift ideas, and prompts the user to follow up, such as asking how a partner’s presentation went.}
\footnotetext{We resized the original sketches for readability.}

Although participants generated a wide range of ideas (\tabref{generated-ideas}, \figref{generated-idea-examples-vertical}), many converged on systems that sense everyday interactions and offer support at specific moments, suggesting that couples envision daily-life sensing and timely intervention as a natural form of attunement support. Despite the variety, we identified some common needs. The first was resolving conflicts and misunderstandings. Three couples (c01, c03, and c04) included conflict resolution as a system feature. They also emphasized the importance of calming down first and included features such as detecting conflict onsets and providing support to help calm down: \textit{``[...] a lot of issues could be solved if you both people are just, like, calmer, and not, like...heated.''} (p07). Regarding conflict resolution, three couples (c01, c03, and c04) also included enhancing emotional understanding as a system feature, with structured prompts to facilitate perspective-taking. The second major need was to objectively assess communication. Two couples (c01 and c05) provided feedback on tone as a feature, and many had positive views on receiving feedback on their conversational tone: \textit{``[...] having an objective, being like, oh wow, if your tone just changed a lot, or your volume got higher, like, I think that's a cool idea to be like, wait, take a step back''} (p03). However, c03 mentioned that they can detect when a conversation drifts off and that they would not need technology support to assess communication. 

\begin{table*}[tbp]
\caption{\tablabel{generated-ideas}Generated ideas from participants during the co-design activities.}
\centerline{\resizebox{1.0\linewidth}{!}{\begin{tabular}{l p{20cm}}
\toprule
\textbf{Couples} & \textbf{Generated ideas} \\
\midrule
c01 & A smart speaker system that detects conflict in at-home conversations and responds with a musical intervention, singing along to playful lines to help calm the moment and introduce some lightheartedness. \\
c02 & A mobile application that tracks partners' interests and schedules using text-based and in-person conversations, and sends reminders and suggestions regarding potential gifts, events of mutual interest, and important personal milestones such as deadlines and exams. \\
c03 & A mobile application that guides couples through difficult conversations using a set of board games that adaptively progress from easier to more difficult questions on a given topic, branching based on how partners rate each exchange. \\
c04 & A mobile application that integrates with messaging applications to detect the onset of conflict and intervene, providing LLM-powered structured prompts that guide conflict resolution and facilitate perspective-taking. \\
c05 & A customizable VR space that offers games designed to enhance collaboration and attunement skills, along with virtual pets that couples can care for together. \\
\bottomrule
\end{tabular}}}
\end{table*}


\subsection{What are couples’ preferences for attunement-enhancing technology? (\textbf{RQ2})}
\seclabel{results-rq2}


\subsubsection{Acceptability of AI-generated feedback}
\seclabel{acceptability-ai-feedback}
We observed significant resistance to receiving AI-generated feedback on couples' communication, especially from LLMs. Some participants expressed concerns about the accuracy and validity of LLM-generated advice, given its limitations in understanding the complexity of human-to-human communication: \textit{``So, like, a lot of the advice could be...could not...like, maybe wouldn't necessarily align with what we're thinking.'' } (p07). p07 and p09 also noted differences in couples' communication styles that LLMs may fail to capture: \textit{``[...] I'm pretty sure if an AI read our text, it would be like, you guys sound so rude to each other.''} (p09). Some participants expressed more general resistance to AI and its feedback on communication: \textit{``[...] I see it very, like, scientifically, because I am in that space, that I'm like, this isn't what it's for.''} (p03);
\textit{``[...] I'm not going to trust what a computer is telling me how to feel.''} (p06). Some participants (p06, p07, and p10) also compared AI with humans and expressed a preference for human feedback over AI for communication assistance: \textit{``But if I were to know, like, for a fact that it was...generated by people and, like, actual...real people, in humans, I think it would...it would actually help me, and I would take it, seriously.''} (p06). Beyond skepticism about AI-generated advice, some participants had concerns about the authenticity of communication. \textit{``[...] it's more like a way of you, like, hiding your emotions [...]''} (p08). One participant also mentioned that receiving feedback may make her hyper-aware of her ``performance'': \textit{``I think sometimes when you introduce technology, it's like, when I'm even focused less on the person, now I'm focused on [...] I'm doing a good job.''} (p03).

\subsubsection{Preference for subtle, just-in-time mediation over advice}
\seclabel{mediator}

Regarding participants' resistance to AI-generated advice, many preferred using AI as a mediator or a nudge rather than as a therapist or a source of advice: \textit{``[...] instead of that thing giving you feedback, it's like a mediator to help you understand the other person.''} (p07). Some participants preferred the second example idea over the first because it felt more like nudging instead of advice: \textit{``Because it's less like...I guess it's more of a suggestion, rather than sounding like a statement.''} (p10). Three couples (c01, c03, and c04) generated ideas aligned with this concept. c01 proposed a system that detects the onset of in-person conflict and nudges users to calm down with music; c03 proposed a board-game app that prompts hard-to-answer questions incrementally to facilitate difficult conversations; c04 proposed an app that detects the onset of text-based conflict and presents a set of guiding prompts to help navigate it. 

\subsubsection{Privacy and control over everyday sensing data}
\seclabel{data-privacy}

Participants held differing views on data privacy. Some expressed strong concerns about sharing personal information, such as conversation content:\textit{``[...] but as far as, like my personal life or emotions, I don't think I would feel comfortable.''} (p05). In contrast, some participants were not very concerned, even about personal information: \textit{``Honestly, like, companies already have so much of our data that I'm just like, yeah, whatever.''} (p10);\textit{``I mean, like, obviously I would want, like, basic privacy protections [...] I wouldn't really mind...''} (p07). We also identified mitigations for privacy concerns. Some participants noted that perceived value would affect their willingness to share: \textit{``maybe if there was a specific problem, or a specific reason to use it in the relationship, I'd be more...open to it [...]''} (p04). Some participants said they would be more comfortable if their data stayed on their phones and was not used for training. Participants also had different comfort levels for each type of data (e.g., tones, text conversations, and in-person conversations), and many emphasized the importance of giving them control over what data to share. 

\subsubsection{Alignment of interface, content, and timing}
\seclabel{interface-timing}
Three couples (c01, c03, and c04) designed technologies for use during conversations, highlighting the need for in-situ sensing and intervention. Meanwhile, our interviews revealed that the acceptability of such support depends on how well the interface, content, and timing align with the interaction context. For in-person conversations, participants wanted interventions that require minimal attention. When we presented the real-time earbud-based cueing system, many participants worried about being distracted from their partner: \textit{``[...] it would be more distracting than beneficial as far as, like, being present with your partner''} (p05). The choice of interface also mattered beyond distraction. Some participants noted that although the earbud system could deliver a subtle nudge without requiring them to look at a screen, they felt uncomfortable wearing earbuds while talking with their partner. Instead, for in-person use, couples generally preferred shared interfaces that both partners could interact with, as reflected in c01's and c04's ideas. In text conversations, participants were more open to longer, richer interactions with the system, suggesting that the acceptable depth of intervention varies by communication channel. Finally, participants emphasized that the technology should fit seamlessly into their daily lives, without requiring them to put on a device before receiving support.

\subsubsection{Bringing fun into attunement enhancement}
\seclabel{bringing-fun}

Almost all participants preferred a gamified approach and appraised our VR gaming idea positively: \textit{``I definitely think that this could be cool, though, and I feel like it's a fun way to, like, reconnect with your partner''} (p05). Some participants preferred approaches that were not explicitly framed as attunement-enhancing but that unconsciously improve their skills: \textit{``[...] when something is like, this will improve your relationship, maybe it, like, feels...personally invasive, [...] that's why, if it was framed differently...''} (p05). Overall, many participants adopted strategies to bring lightheartedness to conflict or difficult conversations: \textit{``[...] every time we got a fight, he just hug me, [...] I think the machine can do something funny, right?''} (p02); \textit{``It helps kinda...bring a little light to a difficult topic for people [...]''} (p05). Beyond difficult conversations, all couples incorporated fun, gamified elements with positive feedback on their progress into their ideas. However, p10 suggested that the preference for a gamified approach may depend on couples' situations: \textit{``[...] if your relationship's in a bad spot, I mean, you're not gonna get to this one (VR games). You should be using one of those things (first and second example ideas).''} (p10).

\subsection{Illustrative example of probing}
\seclabel{appendix-illustrative-example}

We present c04 as an example of how our elicitation technique and co-design activities informed one another. The couple reported miscommunication over text but few difficulties face-to-face. Their questionnaire responses showed the largest self-partner discrepancy in relaxed awareness, so we probed situations in which they wanted their partner to better notice their feelings. p08 described sometimes lacking the bandwidth to spend time with his partner and wishing she would recognize his situation; both agreed that understanding each other’s circumstances was important. When asked how technology might reduce text-based misunderstandings, they initially suggested emojis and pictures. Further probing of how technology could support awareness of a partner’s situation led them to propose a chat-based assistant for navigating conflict.

During co-design, they reacted negatively to LLM-based reflection and earbud cueing tools, concerned LLMs might misinterpret emotions or make communication feel inauthentic. Building on the interview and examples, they designed an LLM-based messaging assistant that detects conflict onset and provides structured prompts. Importantly, they envisioned it as a mediator rather than an advisor, supporting de-escalation, reflection, perspective-taking, problem-solving, and resolution rather than prescribing solutions.

This example illustrates how dyadic discrepancy probing surfaced a couple-specific theme that informed subsequent ideation, while the example technologies helped clarify the couple’s preferred role for AI as a “mediator” rather than a source of advice.

\section{Discussion} 
\seclabel{discussion}

\subsection{Design implications for ubiquitous sensing and intervention for couples}
\seclabel{design-guidelines}

Our findings revealed both cross-cutting and couple-specific preferences. While several high-level preferences recurred across couples, concrete instantiation varied: couples often overlapped on one design dimension while diverging on others, such as addressing similar needs in different contexts or using similar interfaces for different needs. Identifying these couple-specific needs is itself a design task, and our exploratory dyadic discrepancy probe suggests that comparing partners’ perceptions may help surface them. \tabref{design-patterns} illustrates this variation by translating participant-generated concepts into example sensing and intervention configurations, while this section synthesizes the cross-cutting themes.

\begin{table*}[tbp]
\caption{\tablabel{design-patterns}Illustrative sensing and intervention configurations grounded in participant-generated concepts.\protect\footnotemark}
\centerline{\resizebox{1.0\linewidth}{!}{\begin{tabular}{
L{3.4cm}
L{3.4cm}
L{3.0cm}
L{4.0cm}
L{5.8cm}
}
\toprule
\textbf{Support scenario}
&
\textbf{Sensing / user input}
&
\textbf{Trigger / timing}
&
\textbf{Intervention form}
&
\textbf{Safeguard / control}
\\
\midrule

In-person conflict de-escalation (c01)
&
Changes in conversational tone and volume
&
Early signs of escalation
&
Shared cue or music that both partners can hear
&
Mutual opt-in; on-device processing; option to pause sensing; couple-specific calibration
\\

Text-based conflict navigation (c04)
&
Message content and interaction patterns
&
During a potentially escalating exchange
&
Reflective prompts integrated into the messaging interface
&
Mutual opt-in; on-device processing; prompts can be dismissed; couple-specific calibration
\\

Difficult conversation scaffolding in person (c03)
&
Survey and feedback on their situations/emotions
&
Couple-initiated; topics adapt after each input
&
Shared game-like questions that gradually increase in depth
&
Choice of input to provide
\\

Everyday caring and reminders (c02)
&
Relevant conversational and calendar cues
&
Later, when the information becomes relevant
&
Private reminders or suggestions for supporting the partner
&
Choice of input to provide; on-device processing; option to pause sensing; allows review and deletion of stored cues
\\

Playful attunement practice (c05)
&
Interaction cues during shared gaming activities
&
During couple-initiated practice
&
VR games with feedback on coordination and attunement
&
Choice of input to provide
\\

\bottomrule
\end{tabular}}}
\end{table*}

\subsubsection{Support the conversation, do not give advice.}
With the emergence of LLMs, analyzing and intervening in couples' communication has become easier, and some recent systems provide relational advice to users~\cite{chun_conflictlens_2025, fan_is_2026} or even negotiate with their partners on their behalf~\cite{zhang_two_2026}. However, our participants viewed relationship maintenance as deeply human work and strongly resisted receiving AI-generated relational advice. Consistent with prior findings~\cite{wang_futuring_2026, fan_is_2026}, participants felt that AI-generated advice could threaten authenticity and autonomy. Unlike individual AI support, relational AI becomes part of an interaction between two people: its suggestions can influence what one partner says to the other and whether that response is perceived as the partner's own. Systems must therefore support both partners' agency without displacing their direct communication. Future systems should therefore consider more modest roles. One option is to provide analytical insights and let couples draw their own conclusions, as demonstrated by Won et al.~\cite{won_venus_2026}. Another is to offer nudges and mediation that facilitate the couple's own conversation, as explored by Jiang et al.~\cite{jiang_scaffolded_2026}, drawing on the concept of ``mindless computing''~\cite{adams_mindless_2015}. As illustrated in \tabref{design-patterns}, mediation or nudging may take different forms across contexts, ranging from brief shared cues to reflective prompts.

\subsubsection{Calibrate sensing and timing to each couple.}
Participants doubted that AI could understand the complexity and diversity of couples' interactions, for example, worrying that their playful teasing would be misinterpreted as hostility. At the same time, many still expected the system to sense the right moment to intervene. Meeting both expectations requires calibration for each couple. Systems should learn each couple's unique interaction style or allow them to configure it themselves. Personalizing intervention timing has been actively studied for individual interventions~\cite{orzikulova_time2stop_2024, liao_personalized_2020}, but personalization at the dyad level remains largely unexplored and is a concrete modeling challenge for the ubicomp community. 

\footnotetext{These are illustrative examples of design patterns and preferences from our five couples and are not intended to provide generalizable guidelines.}

\subsubsection{Give control over their data.}
Participants' comfort with sharing data varied widely, both across data types (e.g., tone, text, and in-person audio) and among individuals. In a dyadic context, privacy becomes more complex because conversational data involve both partners, whose preferences may differ. Systems should therefore support mutual consent and control over what is sensed, when, and how shared data are used, while clearly communicating the trade-off between data collection and the value of support offered. Depending on the application, this may include selecting which conversations to analyze, temporarily pausing sensing, retaining data on-device, or reviewing and deleting stored cues (\tabref{design-patterns}).

\subsubsection{Design for in-person, in-the-moment support.}
Most current systems focus on text-based interaction~\cite{zhang_two_2026, fan_is_2026, jiang_remini_2025, jiang_scaffolded_2026}, yet our participants expressed a need for in-situ support during in-person conversations. This setting is far less forgiving. The interface, content, and timing must align so the system does not distract partners from each other or feel invasive. As shown in \tabref{design-patterns}, participants also preferred shared interfaces to devices worn or used by a single partner in in-person settings, such as smart speakers and shared mobile applications. How to deliver brief, unobtrusive support during face-to-face interaction is an open interaction design problem well suited to this community.

\subsubsection{Bring fun into hard moments.}
Participants preferred gamified approaches. Gamification is well studied for its effects on engagement in online programs~\cite{looyestyn_does_2017} and in health interventions~\cite{johnson_gamification_2016}. Playful systems for couples also exist~\cite{liu_significant_2021, wang_designing_2023, zheng_pocketbot_2021}. However, these systems primarily aim to enhance connection, whereas our participants desired lightness, especially during conflict and difficult conversations (\tabref{design-patterns}). Designing playful support for a couple's hardest moments, without trivializing them, remains unexplored.

\subsection{Limitations and future work}
\seclabel{limitations}
Our sample size is limited, and the implications presented here are not intended to be broadly generalizable. Rather, we offer them as illustrative starting points that highlight opportunities and challenges for systems supporting couples’ social well-being. Given the limited sample size, we also did not systematically evaluate the dyadic probing technique itself; future work should more systematically assess its effectiveness and utility. Our recruitment strategies likely skewed the sample toward couples with high relationship satisfaction, as those already in a good relationship may be more willing to participate together. We also targeted college students, many of whom were unmarried. Couples' satisfaction and demographics (e.g., age, ethnicity, marital status, and familiarity with technology) are likely to influence needs and preferences; future work should include a broader population. Furthermore, because of the example technology ideas we provided, many findings are explicitly related to AI and VR. Future studies could consider a broader range of technologies and interfaces. Finally, because participants evaluated hypothetical designs, their stated preferences may not fully reflect what they would adopt after hands-on experience. Future work should deploy working prototypes in the field to test whether stated acceptability holds in actual use.

\section{Conclusion} 
\seclabel{conclusion}
We conducted needs-assessment interviews and co-design sessions with five co-located couples to explore technology support for everyday attunement. Couples' designs favored systems that sense interactions, such as conflict onset, and offer brief, subtle support while avoiding AI-generated advice that threatened authenticity and autonomy. We derive preliminary design implications for ubiquitous sensing and intervention, including couple-specific calibration, data control, and support across in-person and text-based interactions. Despite the small sample, this work offers an initial account of what couples may accept from relationship-support technologies and highlights challenges, including dyad-level personalization and unobtrusive in-person intervention. These findings provide a foundation for larger studies of technologies supporting couples' social well-being.		


\bibliographystyle{ACM-Reference-Format}

\bibliography{bibs/local}

\end{document}